\documentstyle{article}
\def\1ad{\mbox{\normalsize $^1$}}
\def\2ad{\mbox{\normalsize $^2$}}
\def\3ad{\mbox{\normalsize $^3$}}
\def\4ad{\mbox{\normalsize $^4$}}
\def\5ad{\mbox{\normalsize $^5$}}
\def\6ad{\mbox{\normalsize $^6$}}
\def\7ad{\mbox{\normalsize $^7$}}
\def\8ad{\mbox{\normalsize $^8$}}
\def\makefront{\vspace*{1cm}\begin{center}
\def\newtitleline{\\ \vskip 5pt}
{\Large\bf\titleline}\\
\vskip 1truecm
{\large\bf\authors}\\
\vskip 5truemm
\addresses
\end{center}
\vskip 1truecm
{\bf Abstract:}
\abstracttext
\vskip 1truecm}
\begin {document}
\def\titleline{
Fine-Tuning of the Cosmological Constant
\newtitleline
in Brane Worlds
}
\def\authors{
Stefan F\"orste
}
\def\addresses{
Physikalisches Institut, Universit\"at Bonn\\
Nu\ss allee 12, D-53115 Bonn, Germany
}
\def\abstracttext{
We discuss how the fine-tuning of the cosmological constant enters brane
world setups. After presenting the Randall Sundrum  
model as a prototype case, we focus on single brane models with
curvature singularities 
which are separated from the brane in the additional dimension.
Finally, the issue of the existence of nearby curved solutions is
addressed. 
}
\large
\makefront

First of all, I would like to mention that this talk is based on
results obtained in collaboration with Zygmunt Lalak, St\'ephane
Lavignac and Hans Peter Nilles\cite{Forste:2000ps,Forste:2000ft}.
One of the most outstanding open problems in theoretical physics is it
to find an explanation for the observed value of the cosmological
constant. As we will discuss below (see also the review
articles\cite{Weinberg:1989cp,Witten:2000zk, Binetruy:2000mh}) a
simple quantum field theoretic estimate provides naturally a
cosmological constant which is at least 60 orders of magnitude to
large. Quantum fluctuations create a vacuum energy which in turn
curves the space much stronger than it is observed. Hence, the
classical vacuum energy needs to be adjusted in a very accurate way in
order to cancel the contributions from quantum effects. The required degree of
accuracy is 60 digits. This is a rather unsatisfactory situation and
theorists are looking for alternatives leading to the observed
cosmological constant in a more natural way. In this talk we will
focus on brane world scenarios, and how they modify the above
mentioned problem. In brane worlds the observed matter is confined to
live on a hypersurface of some higher dimensional space, whereas
gravity and possibly also some other fields can propagate in all
dimensions. This may give some alternative discussion of the
cosmological constant since the vacuum energy generated by quantum
fluctuations of fields living on the brane may not curve the brane
itself but instead the space transverse to it. The idea of brane
worlds dates back to\cite{Rubakov:1983bz,Rubakov:1983bb,Akama:1982jy}.
A concrete realization can be found in the context of string theory
where matter is naturally confined to live on
D-branes\cite{Polchinski:1995mt} or orbifold fixed
planes\cite{Horava:1996qa}. More recently there has been renewed
interest in addressing the problem of the cosmological constant within
brane worlds, for an (incomplete) list of references see
[\ref{braneworlda}--\ref{braneworlde}] and references
therein/thereof.  

The talk is organized as follows. First, we will recall the
cosmological constant problem as it appears in ordinary four
dimensional quantum field theory. The general setup of brane worlds
will be discussed afterwards. We will present a consistency condition
on warped compactifications. Then we will study how fine-tuning 
appears in order to achieve a vanishing cosmological constant in the
Randall Sundrum model\cite{Randall:1999ee,Randall:1999vf}. We will
argue that the solutions presented
in\cite{Arkani-Hamed:2000eg,Kachru:2000hf} are to be fine-tuned once a
singularity is resolved. Finally, we discuss the issue of the
existence of nearby curved solutions.

\vskip0.5cm
\noindent
The observational bound on the cosmological constant is
\begin{equation}\label{observ}
\lambda M_{Pl}^2 \leq 10^{-120} \left( M_{Pl}\right)^4 
\end{equation}
where $M_{Pl}$ is the Planck mass (of about $10^{19}$ GeV) and the
formula has been written such that the quantity appearing on the left
hand side corresponds to the vacuum energy density. In quantum field
theory (including one-loop contributions) this is given by
\begin{equation}\label{qft}
\lambda M_{Pl}^2 = \lambda_0 M_{pl}^2 + \left(\mbox{UV-cutoff}\right)^4
  Str\left( \mbox{\bf 1}\right)
\end{equation}
where $\lambda_0$ is the bare (tree level) value of the cosmological
constant which can be chosen by hand. The supertrace in (\ref{qft}) is
to be taken over degrees of freedom which are light compared to the
scale set by the UV-cutoff. Comparison of (\ref{observ}) with
(\ref{qft}) shows that one needs to fine-tune 120 digits in
$\lambda_0 M_{Pl}^2$ such that it cancels the one-loop contributions
with the necessary accuracy. If one believes that from about a TeV
onwards the world is supersymmetric one still needs to adjust 60
digits. Instead of adjusting input parameters of the theory to such a
high accuracy in order to achieve agreement with observations one would
prefer to get (\ref{observ}) as a prediction or at least as a natural
result of the theory (i.e.\ for example that only a few digits need
to be tuned). The above discussion will be modified in a brane world
setup which we will discuss now. (We should however mention already at
this point that ``modification'' does not stand for improvement of the
situation.)

\vskip0.5cm
In brane worlds matter (charged fermions, gauge fields, etc) is
supposed to be confined to live on a hyper surface (the brane) in a
higher dimensional space, whereas gravity and possibly also some
additional fields can propagate also in directions transverse to the
brane. Since gravitational interactions are much weaker than the other
known interactions, the size of the additional dimension is much
less constrained by observations than in usual
compactifications. Looking for example on product compactifications of
type I string theory it has been noted that it is possible to push the
string scale down to the TeV range when one allows at least two of the
compactified dimensions to be ``extra large'' (i.e.\ up to a $\mu
m$)\cite{Antoniadis:1998ig}. Here, product compactification means
  that the space transverse to the brane is not curved. In particular,
  this implies that any non-vanishing vacuum energy on the brane will
  leed to non-zero curvature on the brane. Hence, product
  compactifications do not give any new view on the cosmological
  constant problem.  
In order to modify the problem one should consider situations where
the vacuum energy on the brane curves the transverse space, and leaves
the brane (almost) flat. This implies that one is looking for warped
compactifications. In the following we will be considering the special
case that the brane is 1+3 dimensional and we have one additional
direction called $y$. Then the ansatz for the five dimensional metric
is in general ($M,N = 0,\ldots ,4$ and $\mu,\nu =0, \ldots 3$)
\begin{equation}
ds^2 \equiv G_{MN}dx^Mdx^N = e^{2A\left(
    y\right)}\tilde{g}_{\mu\nu}dx^\mu dx^\nu + dy^2   
\end{equation}
where the brane will be localized at some $y$.
We split
\begin{equation}
\tilde{g}_{\mu\nu} = \bar{g}_{\mu\nu} + h_{\mu\nu}
\end{equation}
into a vacuum value $\bar{g}_{\mu\nu}$ and fluctuations around it
$h_{\mu\nu}$. For the vacuum value we will be interested in maximally
symmetric spaces, i.e. Minkowski space ($M_4$), de Sitter space
($dS_4$), or anti de 
Sitter space ($adS_4$). In particular, we chose coordinates such that
\begin{equation}\label{ansatz}
\bar{g}_{\mu\nu} =
\left\{\begin{array}{lll}
diag\left(-1,1,1,1\right) & \mbox{for:} & M_4\\
diag\left(
  -1,e^{2\sqrt{\bar{\Lambda}}t},e^{2\sqrt{\bar{\Lambda}}t},
      e^{2\sqrt{\bar{\Lambda}}t}\right)  & \mbox{for:} & dS_4\\
diag\left(-e^{2\sqrt{-\bar{\Lambda}}x^3},e^{2\sqrt{-\bar{\Lambda}}x^3},
  e^{2\sqrt{-\bar{\Lambda}}x^3}, 1\right) & \mbox{for:} & adS_4
\end{array}\right.
\end{equation}
That means we are looking for 5d spaces which are foliated with
maximally symmetric four dimensional slices.
Throughout this talk, the five dimensional action will be of the form
\begin{equation}\label{action}
S_5 = \int d^5 x\sqrt{-G}\left[ R-\frac{4}{3}\left(
    \partial\phi\right)^2 - V\left(\phi\right)\right] - \sum_i\int
    d^5x \sqrt{-g} f_i\left( \phi\right)\delta\left( y-y_i\right) .
\end{equation}
We allow for situations where apart from the graviton also a scalar
$\phi$ propagates in the bulk. The positions of the branes involved
are at $y_i$. With lower case $g$ we denote the induced metric on the
brane which for our ansatz is simply
\begin{equation}
g_{\mu\nu} = G_{MN} \delta^M _\mu \delta^N _\nu .
\end{equation}
The corresponding equations of motion read
\begin{eqnarray}
\sqrt{-G}\left[ R_{MN} -\frac{1}{2}G_{MN}R -\frac{4}{3} \partial_M\phi
  \partial_N\phi +\frac{2}{3}\left(\partial\phi\right)^2 G_{MN}
  +\frac{1}{2} V\left(\phi\right) G_{MN}\right] & &\nonumber \\
  \;\;\; +\frac{1}{2}\sqrt{-g}\sum_i f_i
  \delta\left(y-y_i\right)g_{\mu\nu}\delta^\mu _M\delta^{\nu}_{N} = 0
  & &,
\end{eqnarray}
\begin{equation}
-\frac{\partial V}{\partial \phi}\sqrt{-G}
 +\frac{8}{3}\partial_M\left( \sqrt{-G}G^{MN}\partial_N\phi\right) -
 \sqrt{-g} \sum_i \delta\left( y- y_i\right)\partial_\phi
 f_i\left(\phi\right) = 0 .
\end{equation}
After integrating over the fifth coordinate in (\ref{action}) one
obtains a four dimensional  effective theory. In particular, the
gravity part will be of the form  
\begin{equation}\label{effact}
S_{4,grav} = M_{Pl}^2 \int d^4 x\sqrt{-\tilde{g}}\left(\tilde{R}
  -\lambda\right) , 
\end{equation}
where $\tilde{R}$ is the 4d scalar curvature computed from
$\tilde{g}$. The effective Planck mass is $M_{Pl}^2 = \int dy e^{2A}$,
(note that we put the five dimensional Planck mass to one).
Now, for consistency the ansatz (\ref{ansatz}) should be a stationary
point of (\ref{effact}). This leads to the requirement $\lambda =
6\bar{\Lambda}$. Finally, the on-shell values of the 4d effective
action should be equal to the 5 dimensional one. This results in the
consistency condition\cite{Forste:2000ft} (see also\cite{Ellwanger:2000pq}),
\begin{equation}\label{consistent}
\frac{\left< S_5\right>}{\int d^4 x} = 6\bar{\Lambda}M_{Pl}^2 .  
\end{equation} 
Especially for foliations with Poincare
invariant slices the vacuum energy should vanish.

\vskip0.5cm
As a first example for a warped compactification we want to study the
model presented in\cite{Randall:1999ee}. There is no bulk scalar in
that model. Therefore, we put $\phi = const$ in
(\ref{action}). Moreover, we plug in 
\begin{equation}
V\left( \phi\right) = - \Lambda_B \,\,\, ,\,\,\, f_1 = T_1\,\,\, ,
\,\,\, f_2 = T_2 ,
\end{equation}
where $\Lambda_B$, $T_1$ and $T_2$ are constants. There will be two
branes: one at $y=0$ and a second one at $y=y_0$. Denoting with a
prime a derivative with respect to $y$ the $yy$-component of the
Einstein equation gives
\begin{equation}\label{einst}
6\left( A^\prime\right)^2 = -\frac{\Lambda_B}{4}
\end{equation}
Following\cite{Randall:1999ee} we are looking for solutions being
symmetric under $y \rightarrow -y$ and periodic under $y\rightarrow y
+2 y_0$. The solution to (\ref{einst}) is
\begin{equation}\label{rssol}
A= -\left| y\right| \sqrt{ - \frac{\Lambda_B}{24}} ,
\end{equation}
where $\left| y\right| $ denotes the familiar modulus function for
$-y_0 < y < y_0$ and the periodic continuation if $y$ is outside that
interval. The remaining equation to be solved corresponds to the
$\mu\nu$ components of the Einstein equation,
\begin{equation}\label{einsta}
3A^{\prime\prime} = -\frac{T_1}{4}\delta\left( y\right) -
\frac{T_2}{4}\delta\left( y- y_0\right) .
\end{equation}
This equation is solved automatically by (\ref{rssol}) as long as $y$
is neither $0$ nor $y_0$. 
Integrating equation (\ref{einsta}) from $-\epsilon$ to $\epsilon$,
relates the brane tension $T_1$ to the bulk cosmological constant
$\Lambda_B$, 
\begin{equation}\label{ft1}
T_1 = \sqrt{-24\Lambda_B}.
\end{equation}
Integrating around $y_0$ gives
\begin{equation}\label{ft2}
T_2 = -\sqrt{-24\Lambda_B}.
\end{equation}
These relations arise due to $\bar{\Lambda}=0$ in the ansatz and can
be viewed as fine-tuning conditions for the effective cosmological
constant ($\lambda$ in (\ref{effact}))\cite{DeWolfe:2000cp}. Indeed,
one finds that the consistency condition (\ref{consistent}) is
satisfied only when (\ref{rssol}) together with both fine-tuning
conditions (\ref{ft1}) and (\ref{ft2}) are imposed. Since the brane
tension $T_i$ corresponds to the vacuum energy of matter living in the
corresponding brane, the amount of fine-tuning contained in
(\ref{ft1}), (\ref{ft2}) is of the same order as needed in ordinary 4d
quantum field theory discussed in the beginning of this talk. Here,
however an important question is: What happens if the fine-tunings do
not hold? It has been shown in\cite{DeWolfe:2000cp} that in that case
also solutions exist, however with $\bar{\Lambda}\not= 0$. This
closes the argument of interpreting conditions (\ref{ft1}) and
(\ref{ft2}) as fine-tunings of the cosmological constant.

\vskip0.5cm
\noindent
Next, we would like to discuss a model where the situation seems to be
improved, at first sight. We will focus on the solution discussed
in\cite{Arkani-Hamed:2000eg,Kachru:2000hf} (solution II of the second
reference). In this model there is a bulk scalar without a bulk
potential 
\begin{equation}
V\left( \phi\right) =0 .
\end{equation}
In addition we put one brane at $y=0$, and the bulk scalar couples to
the brane via
\begin{equation}\label{coupling}
f_0 \left(\phi\right) = T_0 e^{b\phi}\,\,\, ,\,\,\,\mbox{with:}\,\,\, b=\mp
\frac{4}{3} .
\end{equation}
For $\bar{\Lambda}=0$, the  bulk equations are solved by $A^\prime
=\pm \frac{1}{3}\phi^\prime$, and
\begin{equation}\label{self}
\phi\left( y\right) =\left\{
\begin{array} {l l l}
\pm \frac{3}{4}\log\left| \frac{4}{3}y +c\right| +d & \mbox{for:}& y< 0 \\ 
\pm \frac{3}{4}\log\left| \frac{4}{3}y -c\right| +d & \mbox{for:}& y>0
\end{array}\right. ,
\end{equation}
where $d$ and $c$ are integration constants (they correspond to moduli
in an effective description).
Finally, by integrating the equations of motion around $y=0$ one
obtains the matching condition
\begin{equation}
T_0 = 4e^{\pm \frac{4}{3}d} .
\end{equation}
This means that the matching condition results in an adjustment of an
integration constant rather than a model parameter (like in the
previously discussed example). So, there seems to be no fine-tuning
involved even though we required $\bar{\Lambda} =0$. As long as one
can ensure that contributions to the vacuum energy on the brane couple
universely to the bulk scalar as given in (\ref{coupling}) it looks as
if one can adjust the vev of a modulus such that Poincare invariance
on the brane is not broken. The uniform coupling of the bulk scalar to
any contribution to the vacuum energy on the brane may be problematic
due to scaling anomalies in the theory living on the
brane\cite{Forste:2000ft}. But apart from that we are going to
discuss even more severe problems of the above proposal, now.
In order to have a chance to be in a agreement with four dimensional
gravity, the five dimensional gravitational wave equation should have
normalizable zero modes in the given background. In other words this
means that the effective four dimensional Planck mass should be
finite. For the considered one brane model and $c < 0$ this implies
that $\int_{-\infty} ^{\infty} dy e^{2A\left( y\right)}  $
  should be finite. However, plugging in the solution (\ref{self})
  one finds that this is not the case. The authors
  of\cite{Arkani-Hamed:2000eg,Kachru:2000hf} proposed to solve this
  problem by choosing $c>0$ and cut off the $y$ integration at the
  singularities at $\left|y\right| =c$. This prescription yields a
  finite four dimensional Planck mass. But now checking the
  consistency condition (\ref{consistent}) one finds that it is not
  satisfied anymore. The explanation for this is simple -- the equation
  of motions are not satisfied at the singularities, and hence for
  $c>0$ (\ref{self}) is not a solution to the equations of motion.  
It has been often observed that singularities appear because at those
points effects become important which are neglected in the effective
description of the theory. (The best known example is perhaps $N=2$
supersymmetric Yang-Mills theory where singularities in the moduli
space are due to monopoles or dyons becoming massless
there\cite{Seiberg:1994rs}.) The hope is that a similar mechanism may
save the solution with $c>0$ and provide a solution to the problem of
the cosmological constant. In the following, we will argue that any
such mechanism is likely to lead to fine-tuning conditions. To this
end, we first modify the theory such that the equations of motion are
satisfied everywhere. This can be easily done by adding two more
branes, situated at $\left| y\right|=c$ to the setup. We choose the
coupling of the bulk scalar to these branes as follows,
\begin{equation}
f_\pm \left(\phi\right)= T_{\pm}e^{b_\pm \phi} ,
\end{equation}
where the $\pm$ index refers to the brane at $y = \pm c$. 
These two additional source terms in the action lead to two more
matching conditions whose solution is
\begin{equation}\label{finet}
b_+ =b_- = b = \mp\frac{4}{3}\,\,\,\,\,\, \mbox{and}\,\,\,\,\,\, T_+ =T_- =
-\frac{1}{2}T_0 . 
\end{equation}
The amount of fine-tuning implied by these conditions is again
determined by the deviation of the vacuum energy on the brane from
the observed value. Hence, nothing improved. 
A short calculation shows that (\ref{finet}) is essential for the
consistency condition (\ref{consistent}) to be satisfied.

\vskip0.5cm
\noindent
So far, we focused on a very specific model. But it is easy to see
that the above criticism also applies to other models (for more
examples see\cite{Forste:2000ft}). The reason is always, that the exact
amount of energy carried away from the brane by the bulk scalar needs
to be absorbed somewhere else. Alternatively, it could be able to flow of to
infinity. But --as we have seen-- it may be problematic to
localize gravity to the brane in that situation. In fact, it has been
proven in\cite{Csaki:2000wz} that localization of gravity is possible
only if there is either a fine-tuning like in the Randall-Sundrum
model (between bulk and brane parameters) or there are naked
singularities like in the models of
\cite{Arkani-Hamed:2000eg,Kachru:2000hf}. For the latter case we have
argued that the consistency condition (\ref{consistent}) implies that
the contribution of the singularities needs to provide a fine-tuned
amount of energy. We have seen this explicitly when studying a simple
way of ``resolving'' the singularities. However, one would expect that
any other resolution will lead to the same conclusions.\footnote{In
  Jan de Boer's lecture a very similar argument was given as
  follows. Any stringy resolution of the singularities is expected to
  provide a quantized amount of vacuum energy. Therefore, the
  contribution from the brane at zero needs to be fine-tuned such that
  it can be canceled by those ``quanta''.}

\vskip0.5cm
\noindent
What remains to be studied is the response of the system once the
fine-tuning (which appears after the singularities are somehow
resolved) 
is relaxed. In general, there are also solutions with
$\bar{\Lambda} \not= 0$ \cite{Kachru:2000xs},  and any fixed value of
$\bar{\Lambda}$ 
results in a fine-tuning condition depending on $\bar{\Lambda}$
\cite{Forste:2000ft}. This means, that relaxing the fine-tuning will
in general result in a non-vanishing effective cosmological constant. 
However, in the specific model which we discussed above
($b=\pm\frac{4}{3}$) there do not exist any nearby curved
solutions\cite{ Arkani-Hamed:2000eg,Kachru:2000xs}. Therefore, in this
special model there is no vacuum if the fine-tuning is not exactly
satisfied. So, is the only possible response of the system to a
mismatch in the fine-tuning to flow (somehow) back to the fine-tuned
solution with $\bar{\Lambda}=0$? (The option that $|b|=\frac{4}{3}$
is relaxed smoothly is excluded since the $|b|\not=\frac{4}{3}$
solutions are not smoothly connected to the former
ones\cite{Kachru:2000hf}.) Another option is that the scalar field
gets a bulk potential. This is studied in\cite{Forste:2000ft}
and the outcome is that depending on the value of the bulk potential
at zero the effective cosmological constant is constrained to a certian
non-zero value.
Hence, in the presence of a bulk potential there are no solutions with
$\bar{\Lambda}=0$. Therefore, apart from finding an explanation for
the very special coupling of the scalar to the brane one would need to
fine-tune the bulk potential in the $\left| b\right| =\frac{4}{3}$
case. 

\vskip0.5cm
\noindent
From the above discussion we conclude that brane worlds give a new
view on the problem of the cosmological constant. However, also in
brane world models fine-tuning is needed to achieve agreement with
observations. The fine-tuning is of the same order of magnitude as the
one needed in ordinary four dimensional field theory. 

\vskip0.5cm
\noindent
{\large \bf Acknowledgements}

\smallskip
\noindent
It is a pleasure to thank St\'ephane Lavignac, Zygmunt Lalak, and
Hans Peter Nilles for a very enjoyable collaboration on the subjects
presented in this talk. Further, I would like to express my gratitude
to the organizers of the  
``34th International Symposium Ahrenshoop on the Theory of Elementary
Particles'' for creating a pleasant and stimulating atmosphere during
the meeting.\\
This work is supported by the European Commission RTN
programs HPRN-CT-2000-00131, 00148 and 00152.


\end{document}